\documentclass{elsart}

\usepackage{natbib}
\usepackage{graphicx}

\usepackage{amssymb}
\usepackage{amsmath}

\def\url#1{{\ttfamily\def\/{/\discretionary{}{}{}}#1}}
\def\bibcode#1{}
\def\ellb{\mbox{\boldmath $\ell$}}
\def\thetab{\mbox{\boldmath $\theta$}}
\begin{document}

\begin{frontmatter}
\title{Weak Lensing of the CMB by Large-Scale Structure}
\author[address1]{Alexandre Amblard\thanksref{aaemail}}, 
\author[address2]{Chris Vale\thanksref{cvemail}}, 
\author[address1,address2]{Martin White\thanksref{mwemail}}
\address[address1]{Department of Astronomy, University of California,
Berkeley, CA, 94720}
\address[address2]{Department of Physics, University of California,
Berkeley, CA, 94720}
\thanks[aaemail]{E-mail: amblard@astro.berkeley.edu}
\thanks[cvemail]{E-mail: cvale@astro.berkeley.edu}
\thanks[mwemail]{E-mail: mwhite@astro.berkeley.edu}

\begin{abstract}

Several recent papers have studied lensing of the CMB by large-scale
structures, which probes the projected matter distribution from
$z=10^3$ to $z\simeq 0$.  This interest is motivated in part by
upcoming high resolution, high sensitivity CMB experiments, such as
APEX/SZ, ACT, SPT or Planck, which should be sensitive to lensing.  In
this paper we examine the reconstruction of the large-scale dark
matter distribution from lensed CMB temperature anisotropies.  We go
beyond previous work in using numerical simulations to include higher
order, non-Gaussian effects and find that the convergence and its
power spectrum are biased, with the bias increasing with the angular
resolution. We also study the contamination by the kinetic
Sunyaev-Zel'dovich signal, which is spectrally indistinguishable from
lensed CMB anisotropies, and find that it leads to an overestimate of
the convergence. We finish by estimating the sensitivity of the
previously cited experiments and find that all of them could detect
the lensing effect, but would be biased at around the 10\% level.

\end{abstract}

\end{frontmatter}

\section{Introduction} \label{sec:intro}

Weak gravitational lensing by large-scale structure has recently
become a powerful tool in the cosmologists' toolbox, allowing us to
map the mass distribution in the universe.  Lensing measurements 
using galaxy ellipticities have already begun to constrain cosmological 
parameters and to test our paradigm for hierarchical structure
formation \citep[see][for a recent review]{vWMe03}.  Even as this 
effort ramps up, yet 
another doorway into the weak lensing gold mine will open as a new 
breed of large surveys, such as the Atacama Cosmology Telescope
(ACT\footnote{http://www.hep.upenn.edu/$\sim$angelica/act/act.html}), 
APEX-SZ\footnote{http://bolo.berkeley.edu/apexsz/}, 
Planck\footnote{http://astro.estec.esa.nl/Planck}, and the 
South Pole Telescope (SPT\footnote{http://astro.uchicago.edu/spt/}), 
begin to come on line.  These surveys will probe the millimeter and 
sub-millimeter wavebands with unprecedented power and resolution, 
and thus enable us to map the large-scale distribution of mass 
in the universe by probing the gravity induced deflections of 
cosmic microwave background (CMB) photons as they 
journey from primordial times at $z\sim 10^3$ to the present.
In addition to providing the first ever map of the projected dark
matter over most of cosmological history, such a measurement may 
have the potential to provide us with precision constraints on 
cosmological parameters, such as the neutrino mass and the dark energy 
equation of state \citep{Kaplinghat03}, by measuring the matter power 
spectrum with percent level accuracy.\\\\
Beginning with the pioneering work of \cite{Zaldarriaga99}, a 
considerable effort has been put into developing an accurate and 
sensitive estimator of the lensing effect.  In this paper, we 
seek to further the effort to develop methods necessary 
for the best possible estimation of the projected mass 
distribution from CMB temperature information.  To this end, 
we examine several effects which have not been studied in earlier 
works, both to see how they influence the reconstruction and 
whether they will provide us an opportunity to enhance the 
signal-to-noise of the estimator.  We note that although we will 
concentrate here on the extraction of information on large angular 
scales, reconstruction at the cluster level \citep{Seljak00} is 
another exciting possibility, which we discuss elsewhere \citep{VAW04}.
The current state of the art techniques comprise maximum likelihood 
estimators \citep{Hirata03} and the computationally more tractable 
quadratic estimators \citep{Hu01a}.  These assume that 
both the primary CMB and the large scale structure responsible 
for the lensing are Gaussian random fields, and that noise 
is both Gaussian and uncorrelated with the signal.  The 
estimators are optimized to solve for the lensing effect by looking for the 
non-Gaussianity induced by the mapping of one Gaussian field by 
another.  Unfortunately, only the first of the four assumptions 
listed above is true.  The projected mass distribution is 
non-Gaussian except on large angular scales, and 
the kinetic Sunyaev-Zel'dovich (kSZ) effect 
\citep[][for recent reviews see \citealt{Reph95,Birk99,Carl02}]{SZ72,SZ80a}
is a particularly pernicious source of confusion because it is
non-Gaussian, spatially correlated with many of the structures doing
the lensing, and spectrally indistinguishable from primary CMB
anisotropies. To test the impact of these issues, we create
lensing and kSZ fields by using an N-body simulation to model relevant
structures, and then apply these fields to random realizations of the
CMB (details of our methods are provided in an Appendix).  The maps
that result from this are then used to reconstruct the projected mass
density and the dark matter power spectrum for various experimental
parameters.  We have elected to use the quadratic estimator of
\cite{Hu01a} for our reconstructions, in part because it is
computationally more tractable than the maximum likelihood estimator,
and because the two methods are predicted to be roughly equivalent for
the angular scales and experimental parameters we are considering
\citep{Hirata03}.  More important, we hope to test the validity of the
\emph{assumptions} of the current generation of estimators, and
thereby gain insight into the issues facing any method which is based
on a Gaussian approximation.\\\\  We note that while our ability to remove
foregrounds and point-sources from the CMB maps may ultimately prove
challenging for any reconstruction, we will not focus on these
complications here.  Instead, we will restrict our analysis to the
observationally irreducible complications discussed above and
instrument effects which are roughly consistent with our fiducial
surveys.  Also, we will consider only the CMB \emph{temperature}
anisotropies, both because this simplifies the calculations and
because it is more relevant for the next generation of wide field
instruments with high angular resolution, which will not initially be
polarization sensitive.  This subject remains of interest, however,
and certainly merits continued study. \\\\ The outline of our paper is
as follows.  In Section \ref{sec:estimator} we describe the estimator
of the projected mass distribution used hereafter, and motivate our
choice.  We then use this estimator to reconstruct a Gaussian
projected mass map in Section \ref{sec:gauss}.  In Section
\ref{sec:nongauss} we investigate the effect of non-Gaussianity in the
lensing field, and in Section \ref{sec:ksz} we include contamination
from the kSZ and describe our efforts to mitigate the problem.  The
effects of these contaminants are considered as a function of
instrument resolution in Section \ref{sec:surveys}, where we also
provide an estimate of how well our fiducial surveys will reconstruct
convergence maps and power spectra.  We then summarize and discuss our
results in Section \ref{sec:summary}.  Finally, some details of the
simulations and the estimator are presented in an Appendix.

\section{Optimal Estimators of the Lensing Potential} 
\label{sec:estimator} 

The use of lensing information contained in high resolution CMB
temperature maps as a probe of the projected matter distribution was
first introduced by \cite{Zaldarriaga99}.  They constructed a particular 
quadratic combination of derivatives of the CMB temperature field that, 
when averaged over realizations of the CMB, returned the projected mass
distribution (or convergence, $\kappa$, in dimensionless units.)  
Since this seminal paper, several authors have sought to improve upon 
the statistical analysis, and a notable step forward (resulting in an 
order of magnitude increase in signal-to-noise) was achieved by the 
proposed optimal
quadratic estimator of \cite{Hu01a}.  Still further improvement may be
possible by employing maximum likelihood techniques \citep{Hirata03}.  
As these are computationally more difficult, and have been shown to be roughly
equivalent to the quadratic estimator for the observational parameters
of interest to us, we do not consider them here.  In this section, we 
briefly remind the reader of the effect of weak lensing on the CMB
\citep[see][for a comprehensive review of weak lensing]{BS01} 
and then introduce the quadratic estimator of \cite{Hu01a}.  We will 
use comoving coordinates, adopt units where the speed of light $c = 1$, 
and we will work in the flat sky 
approximation\footnote{We assume a flat sky to simplify the discussion.  
The extension to a full sky presents no obstacle in principle.}.

In the weak lensing limit, gravitational lensing of CMB light rays that
originate at the surface of last scattering simply remaps the primary
temperature field according to \citep{Seljak96}
\begin{equation} \label{eq:remap}
T(\thetab) = \tilde{T} (\thetab ') = \tilde{T} (\thetab - \delta \thetab)
\end{equation}
where we denote vectors with boldface, $T$({\boldmath $\theta $}) 
is the observed temperature at position 
{\boldmath $\theta $}, $\tilde{T}$({\boldmath $\theta $}$'$) is the unlensed 
temperature at position {\boldmath $\theta $}$'$, and 
$\delta${\boldmath $\theta $} is the deflection angle introduced 
by lensing due to inhomogeneities in the gravitational potential $\phi$ 
along the line of sight
\begin{equation} \label{eq:deflect}
\delta \thetab  = {2 \over \chi_s} \int_0^{\chi_s} d \chi (\chi_s - \chi) 
\mbox{\boldmath $\nabla$}_{\perp} \phi
\end{equation}
where $\chi$ is the radial comoving coordinate, $\chi_s$ is the 
comoving distance to the surface of last scattering, and 
{\boldmath $\nabla$}$_{\perp}$ denotes the spatial gradient perpendicular 
to the path of the light ray.

One important effect of this remapping due to lensing is to 
enhance the non-Gaussian power in the damping tail of the CMB 
temperature anisotropy, as measured by the four-point function 
\citep{Zaldarriaga00}.  \cite{Hu01b} showed that there is a 
quadratic estimator that maximizes the signal-to-noise information
available in this four-point function assuming:

\begin{itemize}
\item The lensing potential is Gaussian random field
\item The noise is Gaussian and uncorrelated with the signal
\item We are interested in information on large angular scales
\item The deflection due to lensing is ``small''
\end{itemize}

The assumption that deflections are small allows us to safely expand 
Eq. (\ref{eq:remap}) to linear order, so that
\begin{equation} \label{eq:linear}
T(\thetab) \simeq \tilde{T}(\thetab) - 
\mbox{\boldmath $\nabla$} \tilde{T} \cdot \delta \thetab
\end{equation}
We will discuss the validity of these approximations in detail 
below, but first let us pause to consider what form a reasonable 
estimator should take.  It is obvious from the Taylor expansion of 
Eq. (\ref{eq:linear}) that the observed temperature contains 
information about the lensing field, and it is also clear that any 
estimator chosen must satisfy 
$ \langle \kappa_{\rm est} \rangle_{\rm{CMB}} = \kappa$ 
when averaged over many realizations of the CMB.  Furthermore, 
the estimator must contain an even number of temperature terms, 
since the expectation value is zero for odd powers of temperature.  
The simplest estimator should therefore be proportional to $T^2$, 
normalized so that 
$ \langle \kappa_{\rm est} \rangle_{\rm{CMB}} = \kappa$ 
and filtered to maximize the signal-to-noise, so that 
it takes the form in Fourier space
\begin{equation} \label{eq:hustimator}
\kappa_{\rm est} (\ellb) = {A_\ell \over 2} 
\int {d^2 \ell_1 \over (2 \pi)^2}
F(\ellb_1,\ellb_2) \ T(\ellb_1) T(\ellb_2)
\end{equation}
where {\boldmath $\ell \equiv \ell_1 + \ell_2$}.  
Eq. (\ref{eq:hustimator}) is indeed the optimal quadratic 
estimator of \cite{Hu01a} when 
\begin{equation} \label{eq:filter}
F(\ellb_1,\ellb_2) \equiv {\ellb \cdot (
\ellb_1 \tilde{C_{\ell_1}} + \ellb_2 \tilde{C_{\ell_2}}) 
\over 2 C_{\ell_1}^{\rm tot} C_{\ell_2}^{\rm tot} }
\end{equation}
is a filter that optimizes the signal-to-noise, where 
$\tilde{C_{\ell}}$ is the unlensed primary CMB temperature 
power spectrum, $C_{\ell}^{\rm tot}$ is the measured spectrum 
(including lensing and noise), and 
\begin{equation} \label{eq:N_L}
A_{\ell}^{-1} = {\ell}^2 \int{d^2 {\ell}_1 \over (2\pi)^2} 
2 C_{\ell_1}^{\rm tot} C_{\ell_2}^{\rm tot} F^2 (\ellb_1,\ellb_2)
\end{equation}
is the normalization.  As we shall see below, $A_\ell$ performs 
two roles, since it also serves as the $0^{\rm th}$ order 
noise term (Eq. \ref{eq:noise}) in the estimated power spectrum 
when the assumptions listed above are satisfied.

We note that although the general characteristics of this estimator
are intuitively clear, the original analysis to determine the exact
form of the best estimator is somewhat involved 
\citep{Hu01b,Cooray03}. However \cite{Hirata03}
showed that this estimator (equation \ref{eq:hustimator}) is simply
the standard minimum variance quadratic estimator (\cite{Tegmark} for
a review) for $\kappa$ small enough so that the linear approximation
is valid. Let us now turn our attention to the main focus of this
paper, the reconstruction of convergence maps and power spectra in the
context of numerical simulations.

In the next section we shall consider map and power spectrum
reconstruction using a Gaussian lensing field and uncorrelated 
Gaussian noise, as assumed by
the estimator.  We shall see that a signal-dependent noise bias
is present for high resolution, high sensitivity experiments as
has already been noted by \citep{Cooray03,Kesden03}.
We show that this bias is large for interesting observational parameters.
We then test the effect of using a non-Gaussian mass distribution to
compute the lensing field in Section \ref{sec:nongauss}, and in Section 
\ref{sec:ksz} we consider the impact of adding a foreground 
contaminant, in this case the kSZ, which is correlated with
the lensing signal.  

\section{Reconstruction of a Gaussian Lensing Field} \label{sec:gauss}

We begin our efforts by reconstructing a convergence field using a
simulated temperature field which has been constructed following the
assumptions of the optimal quadratic estimator \citep{Hu01a}.  
Specifically, we assume a Gaussian CMB sky and produce a Gaussian 
$\kappa$ field with which to lens it, and then use the quadratic 
estimator to reconstruct $\kappa$.  Although we are ultimately 
motivated by our fiducial surveys (and return to them in Section 
\ref{sec:surveys}), our intention here and in Section \ref{sec:tests} 
is to investigate the reconstruction technique under 
``ideal'' conditions.  Since the presence of even a small amount of noise 
provides a reasonable high-$\ell$ cutoff in the estimator, we include 
it at a level appropriate for a very deep integration.
For this reason (and to reduce the computational 
requirements; see Section \ref{Nell}) we adopt a fiducial experiment 
with uniform, white instrument noise of $2\,\mu \rm K$-arcmin (much 
better than the stated goals of the surveys) and $0.^\prime 8$ angular 
resolution.  

\begin{figure}[h]
\begin{center}
\includegraphics[width=12cm]{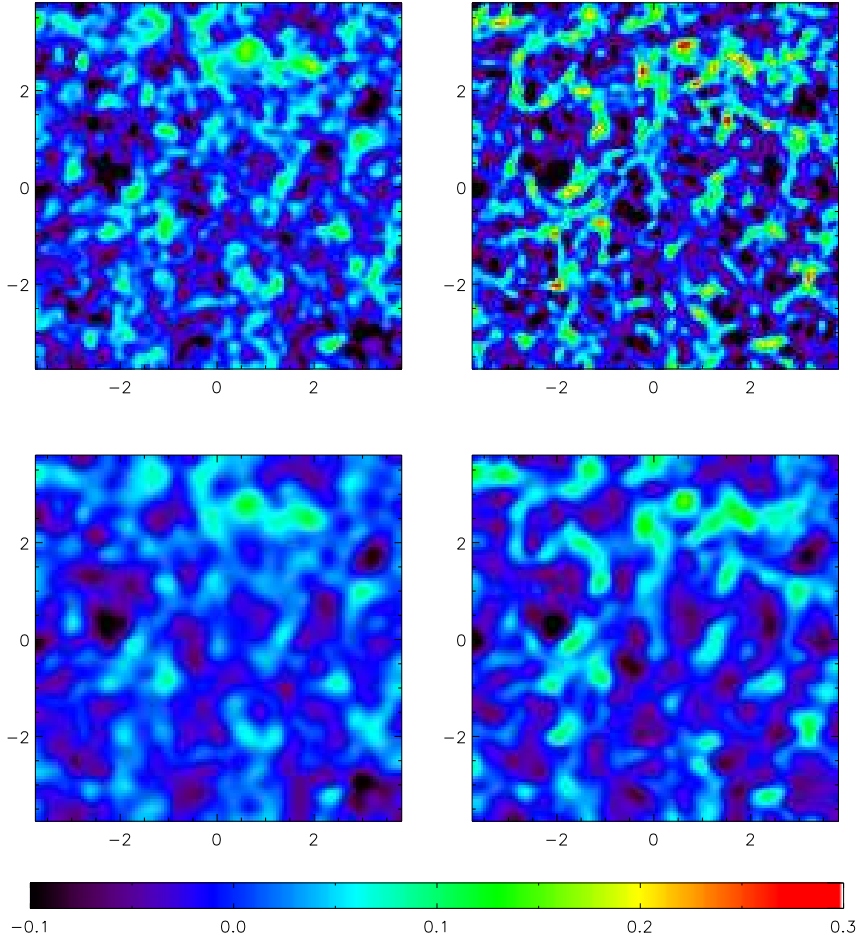}
\end{center}
\caption{Maps of (left) input and (right) reconstructed projected mass, 
in units of 
dimensionless convergence $\kappa$, for two smoothing scales and 
$2 \mu \rm K$-arcmin uncorrelated Gaussian instrument noise.  
While the (top) input and reconstructed 
$\kappa$ maps smoothed by a $10^\prime$ FWHM Gaussian beam do 
contain many of the same structures, the agreement between the two 
(bottom) visibly improves when they are smoothed on a 
$20^\prime$ scale.  The input $\kappa$ maps shown here are Gaussian 
random fields, and all the maps are $7.5^\circ\times 7.5^\circ$. }
\label{fig:Grec}
\end{figure}

An example of an input and reconstructed $\kappa$ map is given in 
Fig. \ref{fig:Grec}.  The raw maps are very noisy, so we have smoothed 
the convergence with a (upper panels) $10^\prime$ and (lower panels) 
$20^\prime$ FWHM Gaussian filter.  Although $10^\prime$ is many 
times the nominal resolution used in the reconstruction, the estimated 
map made at this level of smoothing clearly leaves much to be desired.  
Increasing the smoothing to $20^\prime$ improves the reconstruction 
enough so that many of the features in the maps are reproduced, e.g. 
the overdensities at $(0^\circ,3^\circ) \rm{and} (-1^\circ,-1^\circ)$.  
The reconstruction is still noisy enough that some features, e.g.~at 
$(3^\circ,3^\circ)$, are spurious. 

It seems clear from Fig. \ref{fig:Grec} that if a reconstruction
of the projected mass back to the surface of last scattering 
using only CMB temperature information is to be achieved, it can 
only be on large angular scales.  This means averaging the
density not only over the $10\,$Gpc along the line-of-sight but also 
over $100\,$Mpc transverse to it.  Since we do not expect structures 
on these scales to be particularly informative, we shift our attention 
to the following statistical measures of the convergence field:
\begin{figure}[h]
\centerline{
\includegraphics[width=12cm]{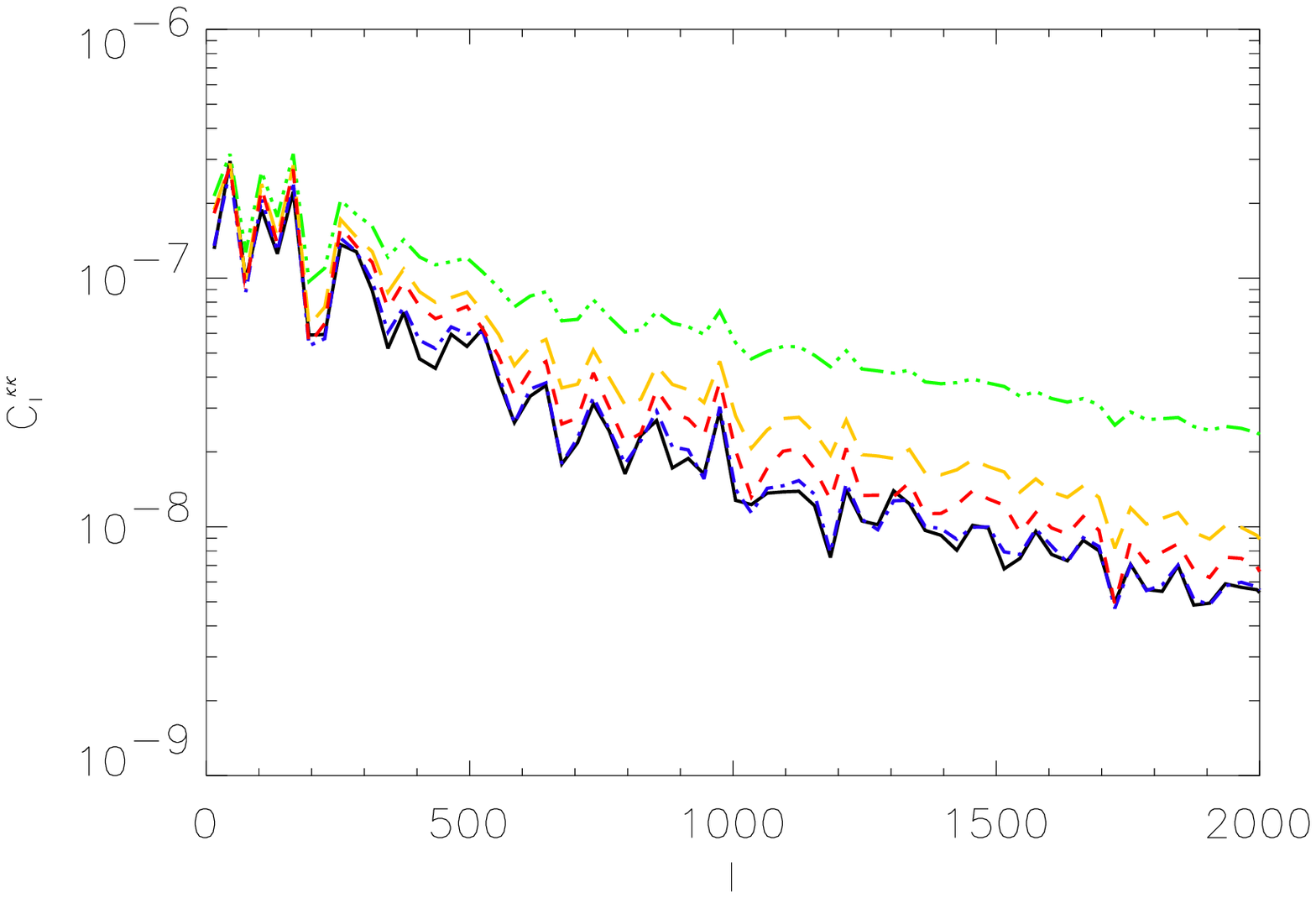} }
{\caption{Reconstruction of the power spectrum for a Gaussian 
convergence map.  The input $C_{\ell}^{\kappa\kappa}$ convergence spectrum 
(black) is poorly reconstructed, as is clear in the auto-spectrum 
(green) if noise terms are not subtracted.  The situation improves 
markedly as the noise is subtracted to first order (orange), and still 
more when second order terms are subtracted (red), while the 
cross-spectrum (blue) fit is quite good. }
\label{fig:Gspec}}
\end{figure}
\begin{figure}[h]
\centerline{
\includegraphics[width=12cm]{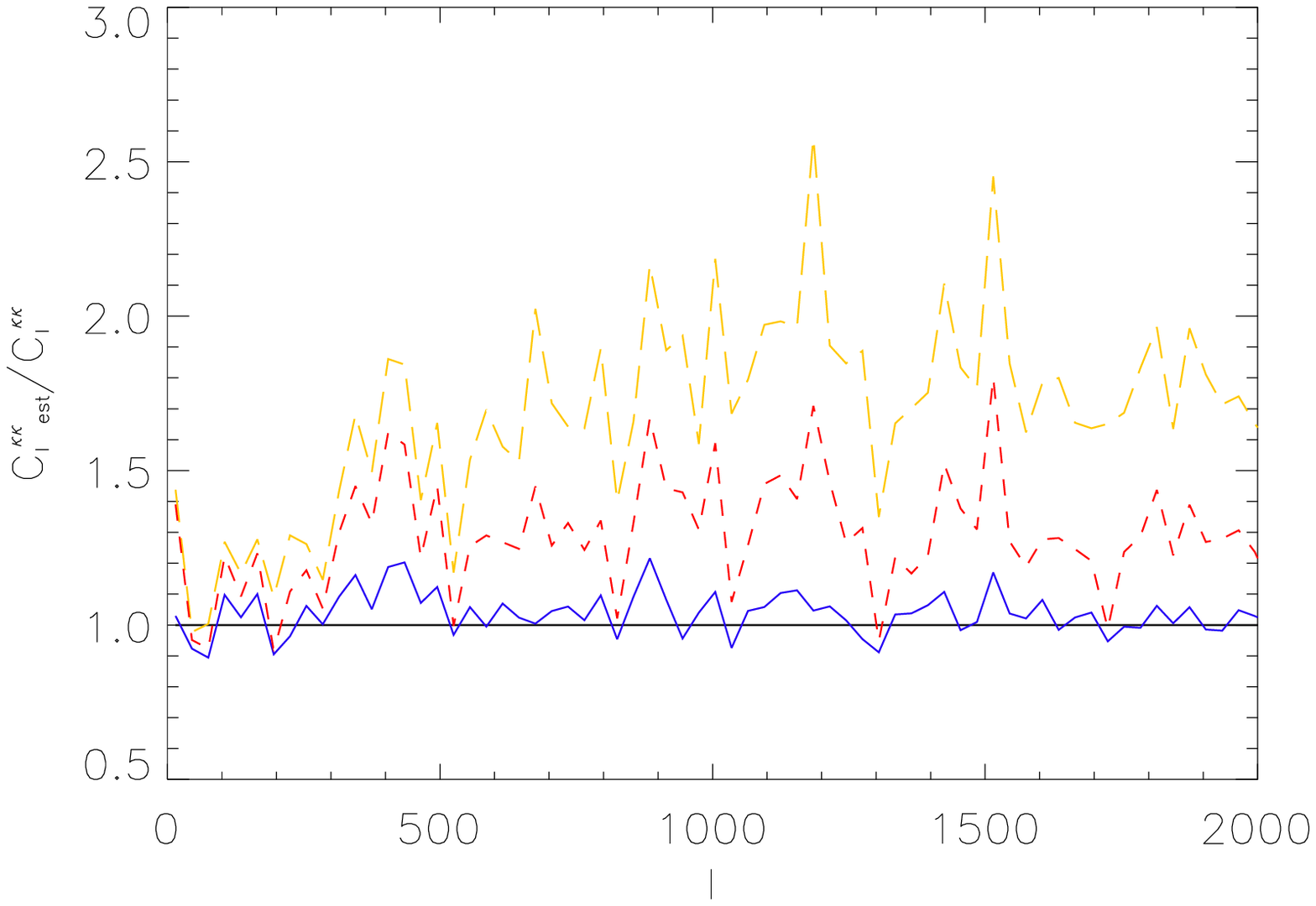} }
{\caption{The ratio of the reconstructed power spectra to the input 
spectrum for a Gaussian $\kappa$ map (the spectra themselves are given 
in Fig. \ref{fig:Gspec}).  The ratio for the cross-spectrum case (blue) 
is consistent with a no multiplicative bias;  however, as is evident 
from the plot of $C_{\ell}^{\rm est}$ (red), there is still a 
substantial ($\sim 25\%$) additive bias even when second order noise 
terms are included.  If only the first order correction is included 
in the noise estimate (Eq. \ref{eq:noise}), the bias is close to 
$\sim 70\%$. }
\label{fig:Gspec2}}
\end{figure}
\begin{eqnarray} \label{eq:spectra}
\nonumber 
\ \ \ \ \ \ \ \ \ \ \ \ \ \ \ \ \ \ \ \ \ \ \ \ \ \ 
\langle \kappa (\ellb) \kappa^* (\ellb^\prime) 
\rangle &&= 2\pi\delta(\ellb - \ellb^\prime)C_{\ell}^{\kappa\kappa} \\
\nonumber 
\langle \kappa_{\rm est} (\ellb) \kappa^*_{\rm est}(\ellb^\prime) 
\rangle &&= 2\pi\delta(\ellb - \ellb^\prime)C_{\ell}^{\rm auto} \\
\langle \kappa_{\rm est} (\ellb) \kappa^* (\ellb^\prime) 
\rangle &&= 2\pi\delta(\ellb - \ellb^\prime)C_{\ell}^{\rm cross} 
\end{eqnarray}
where $C_{\ell}^{\kappa\kappa}$ is the power spectrum of the input 
$\kappa$ map, $C_{\ell}^{\rm auto}$ is the auto-spectrum of the 
estimated map, and $C_{\ell}^{\rm cross}$ is the cross-correlation 
spectrum between the input and estimated convergence.  Of these, 
only $C_{\ell}^{\rm auto}$ is in principal observable, but we will 
make use of all 3 to diagnose the quality of the reconstruction.

We generically expect that any estimator of $\kappa$ will have both 
an uncorrelated noise term $n_\ell $ and a multiplicative bias term 
$b_\ell$ associated with it, so that
\begin{equation} \label{eq:kbias}
\kappa_{\rm est} (\ellb) = b_\ell \kappa (\ellb ) + n_\ell
\end{equation}
If the estimator is properly normalized by $A_\ell$ as defined in 
Eq. \ref{eq:N_L}, then the multiplicative term $b_\ell = 1$, and the 
estimator is unbiased.  On the other hand, $C_{\ell}^{\rm auto}$ is simply 
related to the input power spectrum 
$C_{\ell}^{\kappa\kappa}$ via the noise power $\mathcal{N}_\ell$ as 
$C_{\ell}^{\rm auto} = C_{\ell}^{\kappa\kappa} + \mathcal{N}_\ell$, so 
that our estimate $C_{\ell}^{\rm est}$ of $C_{\ell}^{\kappa\kappa}$ is
\begin{equation} \label{eq:bias}
C_{\ell}^{\rm est} = C_{\ell}^{\rm auto} - \mathcal{N}_{\ell}^{\rm est}
\end{equation}
where $\mathcal{N}_{\ell}^{\rm est}$ is the the estimated noise power. 
Thus, even if the normalization $A_\ell$ is correct, so that multiplicative 
bias is not an issue, an unbiased estimate of $C_{\ell}^{\kappa\kappa}$ 
requires that we know the noise power in the map so that we can 
account for the additive bias $\mathcal{N}_\ell$.  Fortunately, 
\cite{Cooray03} have shown that it is possible to estimate the noise as
\begin{equation} \label{eq:noise}
\mathcal{N}_{\ell}^{\rm est} = 
{\ell^2\over 4}\left( A_\ell + A_\ell^{NG} + \cdots \right)
\end{equation}
where $A_\ell$ is the normalization of \cite{Hu01a}, $A_\ell^{NG}$ 
is an additional signal dependent term (Eq. \ref{eq:ng}), and 
$\cdots$ represents higher order terms.  We will define any 
failure to properly estimate the noise contribution 
$\mathcal{N}_{\ell}$ by $\mathcal{N}_{\ell}^{\rm est}$ 
as an additive bias.

Although $C_{\ell}^{\rm cross}$ is not measurable, we can still make use 
of it as a diagnostic of our estimate.  Specifically, since 
\begin{equation} \label{cross}
\langle C_{\ell}^{\rm cross} \rangle = b_{\ell} C_{\ell}^{\kappa\kappa}
\end{equation}
we will be able to identify a multiplicative bias, caused by a failure of 
$A_\ell$ to properly normalize the estimator, 
and an additive bias, which exists when we are unable to determine the noise 
power in the reconstruction.

We display these power spectra in Fig. \ref{fig:Gspec}, and for clarity, 
we include the same information in the ratio plot shown in Fig. 
\ref{fig:Gspec2}.  It is clear from these plots that the (blue) 
cross spectrum 
$C_{\ell}^{\rm cross}$ traces the (black) convergence power spectrum 
$C_{\ell}^{\kappa\kappa}$, so the quadratic estimator is not 
multiplicatively biased if $A_\ell$ (as defined in Eq. \ref{eq:N_L}) 
is used for the normalization  and the assumptions of the estimator 
are met.  

However, we see in the same figures that using 
$\mathcal{N}_{\ell}^{\rm est} \simeq A_{\ell}$ will lead to a substantial 
additive bias ($\sim 70\%$).  Thus,
quantitatively, an accurate measure of $C_\ell^{\kappa\kappa}$
requires us to use the higher order corrections, which unfortunately
depend on $C_\ell^{\kappa\kappa}$ itself.  We have two options here.
The first is to develop an iterative scheme to solve for
$C_\ell^{\kappa\kappa}$.  This is non-trivial because the estimate of
$C_\ell^{\kappa\kappa}$ is so intrinsically noisy.  Some heavy
smoothing is required to regularize the procedure.  We have instead
decided to assume that a parameterized functional form for
$C_\ell^{\kappa\kappa}$ would be available, along with a good enough
first guess at the parameters that the bias can be accurately
estimated.  Specifically we compute the bias assuming the ``true''
spectrum used in constructing the simulation, and use the residual as
an estimate of how well we do.

The next term in $\mathcal{N}_\ell$ beyond $A_\ell$ is given in
Eq. \ref{eq:ng}, and labeled schematically in Eq. \ref{eq:noise}
as $A_\ell^{NG}$.  This term reduces the additive bias in the
reconstruction to about $25\%$.  As the series appears to be 
converging, it is not unreasonable to expect that including more 
higher order terms 
in the computation of $\mathcal{N}_\ell$ would reduce this discrepancy 
even further.  We have not chosen to pursue this calculation here;  
other factors, such as the effect of non-Gaussianity in the 
$\kappa$ map and the contribution of the kSZ, are shown below to enter 
at a level equal or greater than these higher order ``Gaussian 
approximation'' terms (also, the computational resources required are 
non-trivial; see Section \ref{Nell}).  In the next section, 
we address these effects for our fiducial deep integration survey, 
and then in Section \ref{sec:surveys} we consider what we have 
learned in the context of ACT, APEX-SZ, Planck, and SPT.

\section{Some Complicating Factors} \label{sec:tests}

So far, we have considered the reconstruction of the projected mass 
distribution under the simplifying assumption that the convergence 
and the noise are uncorrelated Gaussian fields.  In this section, we 
explore some complicating factors, first by modifying the 
lensing field in Section \ref{sec:nongauss}, and then by including 
a pernicious foreground, the kSZ, in Section \ref{sec:ksz}.

\subsection{The Effect of a Non-Gaussian $\kappa$ Field} 
\label{sec:nongauss}

\begin{figure}[!h]
\centerline{
\includegraphics[width=12cm]{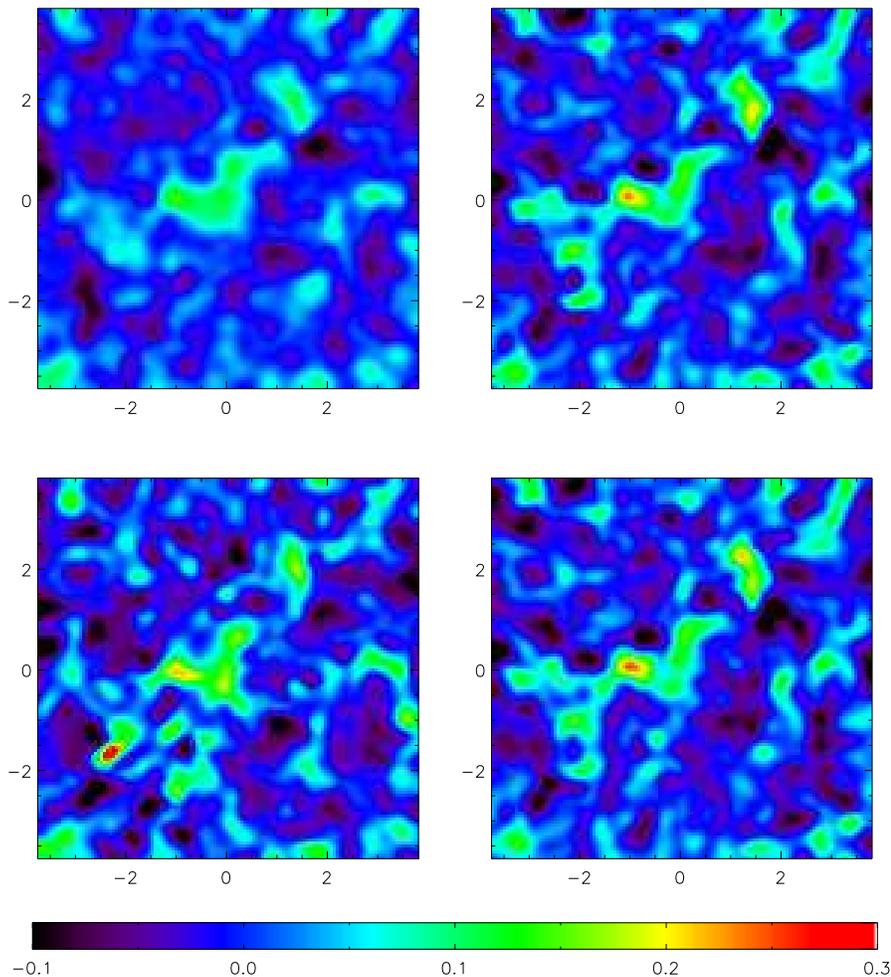}}
\caption{Reconstructions of a (top left) input non-Gaussian $\kappa$ map.  
The (top right) map has been reconstructed in the absence of 
foregrounds, and while it clearly contains much of the same structure 
as the input map, the correspondence is far from perfect, as we discuss 
in Section \ref{sec:nongauss}.  When the kSZ is included, the 
(bottom left) reconstruction still retains some visual similarity to 
the original, although additional spurious features are added to the 
map, such as the hot spot near $(-2^\circ,-2^\circ)$.  In the last panel 
we show the (bottom right) reconstruction when the kSZ is masked using 
the technique described in Section \ref{sec:ksz}.  The worst of the 
spurious structures caused by the kSZ have disappeared, and the map 
bears a strong resemblance to the reconstruction made in the absence 
of kSZ.  The maps are $7.5^\circ\times 7.5^\circ$ fields sampled at 
$0.^{\prime}8$ resolution, include $2 \mu$K-arcmin of instrument 
noise, and have been smoothed by a $20^\prime$ FWHM Gaussian window 
for presentation. }
\label{fig:NGwSZrec}
\end{figure}

In this section we examine the impact of introducing non-Gaussianity 
into the mass distribution.  We do this by creating a convergence map 
using an N-body simulation to model structures at $z < 2$ (structures 
at higher redshift are modeled by a Gaussian random field as before.)  
We show the resulting $\kappa$ map and its reconstruction in the upper 
panels of Fig. \ref{fig:NGwSZrec}.  The maps have been smoothed 
by a $20^\prime$ window, and on this scale many of the prominent 
features in the recovered map look quite similar to the ``real'' 
ones.  We can be more quantitative by considering the difference map 
of the the input and reconstructed fields, which we find is not well 
correlated with the input map.  However, the distribution of this 
``noise'' map is mildly non-Gaussian (slightly skew positive) and also 
not quite consistent with a ``white noise'' approximation.
\begin{figure}[h]
\centerline{
\includegraphics[width=12cm]{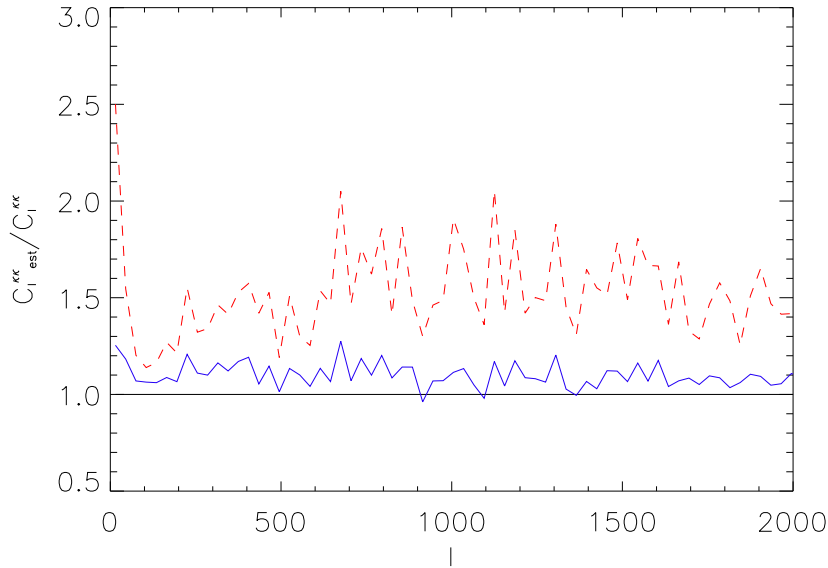}}
{\caption{As Fig. \ref{fig:Gspec2}, but now including the non-Gaussian 
contribution to the $\kappa$ map.  The cross-spectrum (blue) is now 
biased high by roughly $10\%$, and the total bias in the estimated 
convergence spectrum (red) is about $50\%$. }
\label{fig:NGspec}}
\end{figure}

To get a better idea of the properties of the estimated map, we once 
again turn to the convergence power spectrum $C_{\ell}^{\kappa\kappa}$.  
The ratio of the cross spectrum to the power spectrum 
$C_{\ell}^{\rm cross} / C_{\ell}^{\kappa\kappa}$
is shown as the blue (solid) line in Fig. \ref{fig:NGspec}, and indicates 
that the map is now multiplicatively biased high by $\sim 10\%$.  
This bias is not unexpected, because higher order moments of the 
field, which vanished in the Gaussian case, can now contribute
(see Appendix).
The red (dashed) line on the same plot, the ratio of the estimated power 
spectrum $C_{\ell}^{\rm est} / C_{\ell}^{\kappa\kappa}$ (including all the 
corrections considered in Section \ref{sec:gauss}), shows a total 
bias of about $50\%$.

In addition  to the excess power on the scales of 
interest (larger than 6 arcminutes), the reconstructed 
maps tend to have too little power on smaller scales 
(smaller than 2-3 arcminutes, not visible on the map).  
This decrement on very small scales is of similar size 
to the increment on larger scales ($\sim 10\,-\,20\%$).
This is not surprising, given that the quadratic estimator 
has been optimized for a Gaussian field, and other
reconstruction methods based on Gaussian assumptions (e.g. a Wiener
filter) will typically mix power onto large scales from features that
are too sharp to fit the Gaussian approximation.

\subsection{Kinetic SZ Contamination} \label{sec:ksz} 

In this section, we will consider one more wrinkle in the reconstruction:
foreground removal.  While it may be that point sources or diffuse
foregrounds will place real-world limits on how well we can reconstruct
the lensing signal, in principle the only foregrounds we cannot
distinguish from the CMB spectrally are the kinetic Sunyaev-Zel'dovich
effect (kSZ), and its related cousin, the Ostriker-Vishniac (OV) 
effect \citep{OV86}.  Of the two, the kSZ is likely to be more 
troublesome, since it is highly non-Gaussian and correlated to the 
lensing signal.  Therefore, we will address only the kSZ here, although 
we note that this may be somewhat optimistic.

\begin{figure}[h]
\centerline{
\includegraphics[height=5.6cm]{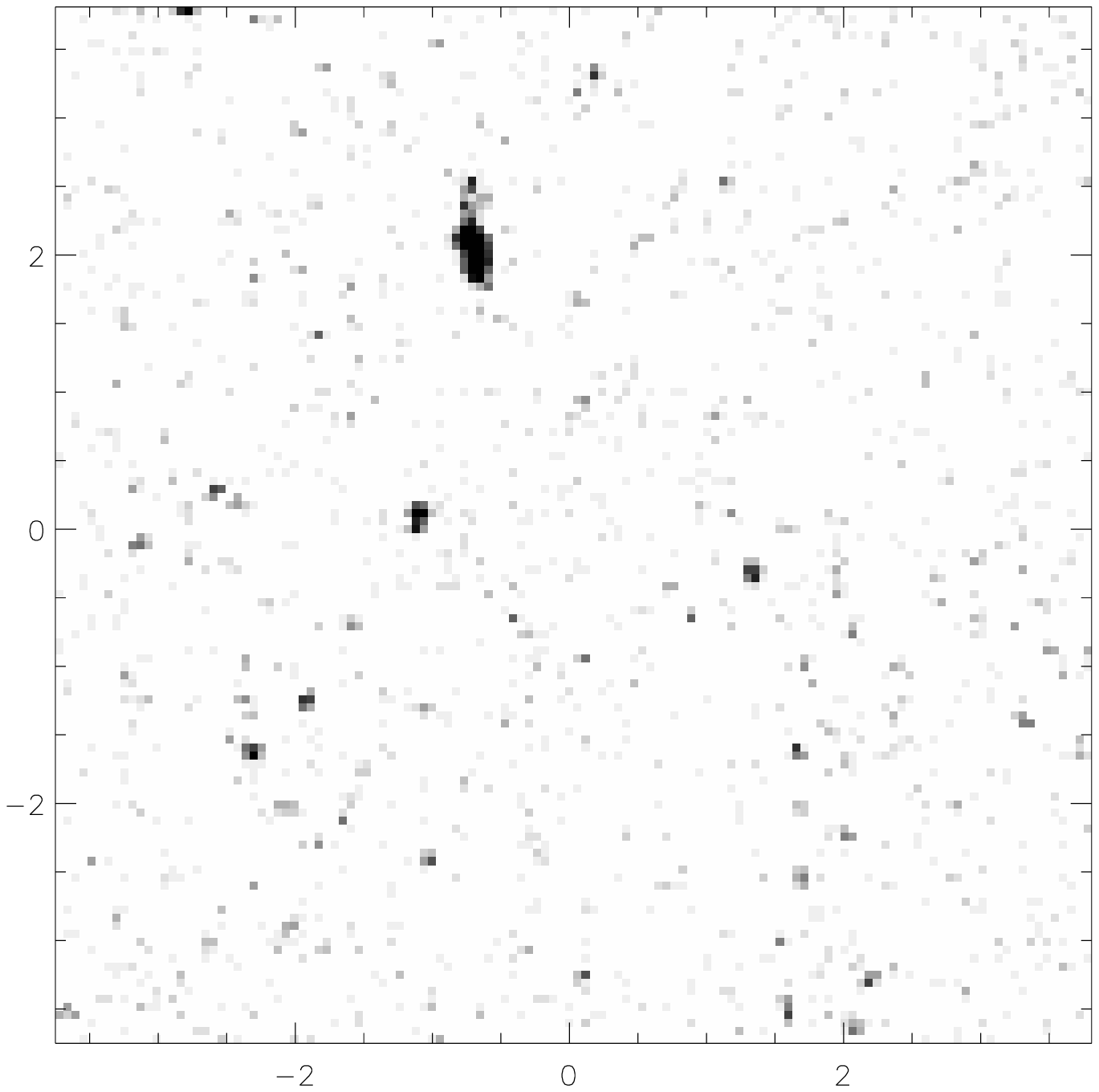}
\includegraphics[height=7.1cm]{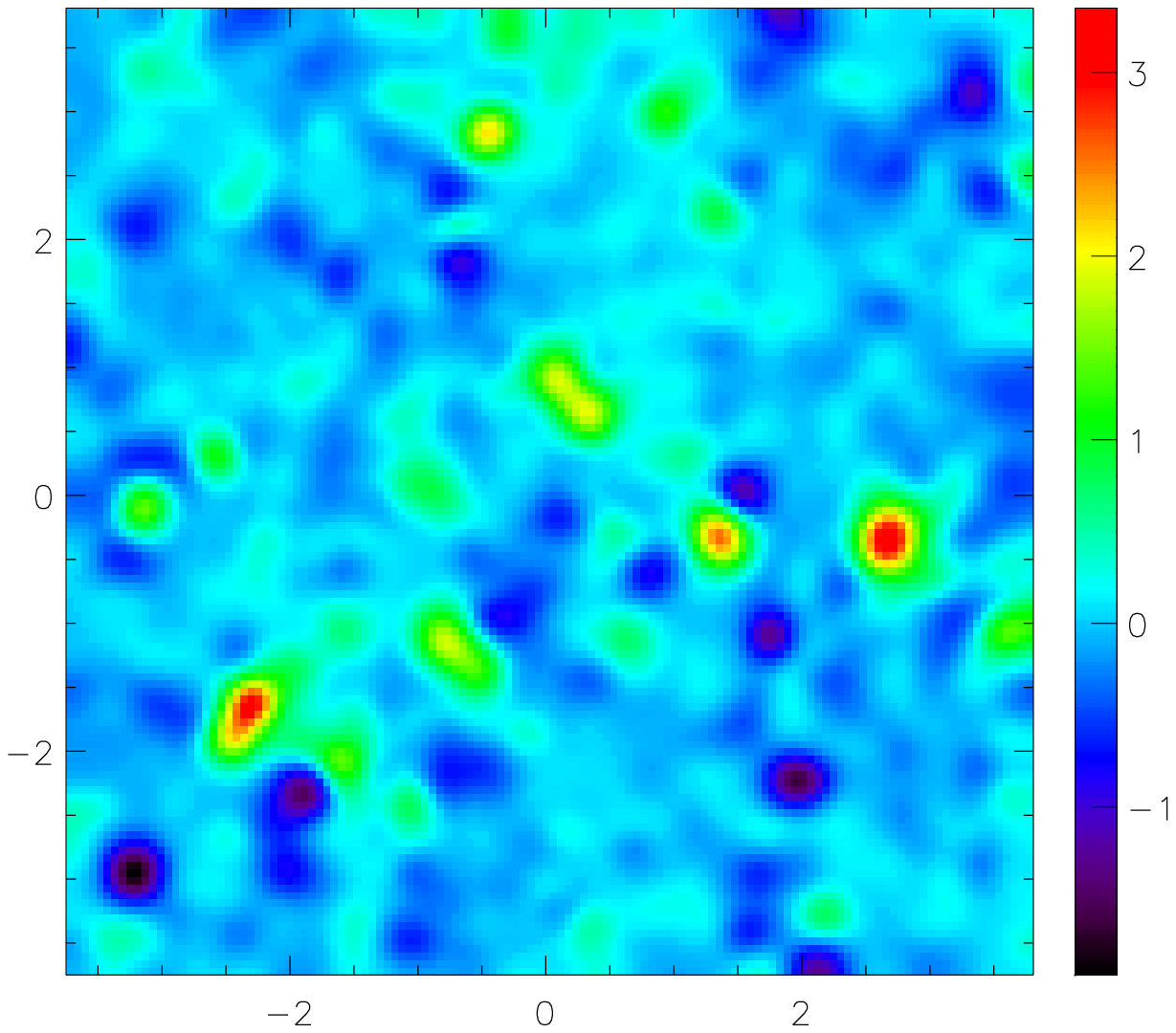}}
{\caption{An illustration of the masking and interpolation techniques 
we use to control the kSZ.  Regions with a thermal SZ temperature 
magnitude in excess of $50 \mu K$ (left) are masked, resulting in a 
loss of only 1.4\% of the pixels in the map.
This can be compared to the (smoothed) kSZ map (right).}
\label{fig:mask}}
\end{figure}

As we discuss in Section \ref{simulations}, we create the kSZ signal 
using the same N-body simulation that we used to make the lensing maps.  
This is then added to the lensed CMB map, which is used to 
reconstruct the convergence $\kappa$.  The resulting map, shown in 
the (bottom left) panel of Fig. \ref{fig:NGwSZrec}, is clearly 
substantially degraded, as the strongest kSZ sources
(Fig. \ref{fig:mask}),  e.g. the 
``hot spot'' at $(-2^\circ,-2^\circ)$, now appear in the reconstructed 
map.  The contamination is even more prominent in the estimated 
convergence power spectrum $C_{\ell}^{\rm est}$, which is now biased 
high by well over a factor of 2!
\begin{figure}[h]
\centerline{
\includegraphics[width=12cm]{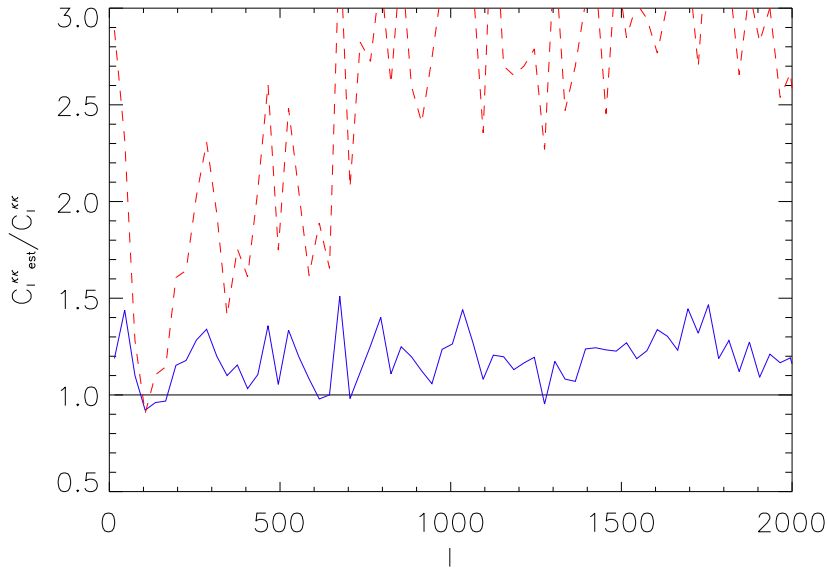}}
{\caption{As Fig. \ref{fig:Gspec2}, but now including effects from both the 
kSZ and the non-Gaussian $\kappa$ field.  The cross spectrum (blue) 
is now biased by 20\%, while for the estimated convergence spectrum 
(red), the bias from the kSZ signal completely overwhelms the signal. }
\label{fig:NGwSZspec}}
\end{figure}

Since the kSZ is concentrated around large clusters, one obvious idea to 
correct for the effect is to ``mask'' the major clusters using a 
measurement of the thermal SZ.  As a simple check of this technique, we 
used a 50$\mu$K (at 150 GHz) thermal SZ temperature threshold to cut 
pixels from the ``observed'' CMB temperature map.  This resulted in our 
excising roughly $1.4\%$ of the pixels from the map, as shown in 
Fig. \ref{fig:mask}.  We then replace the ``holes'' by interpolating 
the surrounding area using the two variable method of 
\cite{Renka84}\footnote{ This method uses a weighted least squares 
approach to reconstruct a surface with continuous first derivatives.}.  
The interpolated map does not 
contain obvious numerical artifacts associated with the masking 
process.

Before we proceed with reconstructions using these masked maps, 
we need to check whether the procedure has significantly reduced the 
lensing signal by limiting the contribution from clusters.  We do 
this by masking and interpolating the lensed temperature map, but 
without adding the kSZ, and then computing the power spectrum.  We 
find that the 1.4\% cut we have used 
results in only a modest (slightly less than 5\%) reduction of the 
lensing signal.  Two issues are likely to be responsible for mitigating 
the masking effect on the lensing power.  First, the lensing signal 
extends to larger angular scales than the kSZ, and second, much of 
the lensing power comes from lower mass halos which we don't mask. 
\begin{figure}[h]
\centerline{
\includegraphics[width=12cm]{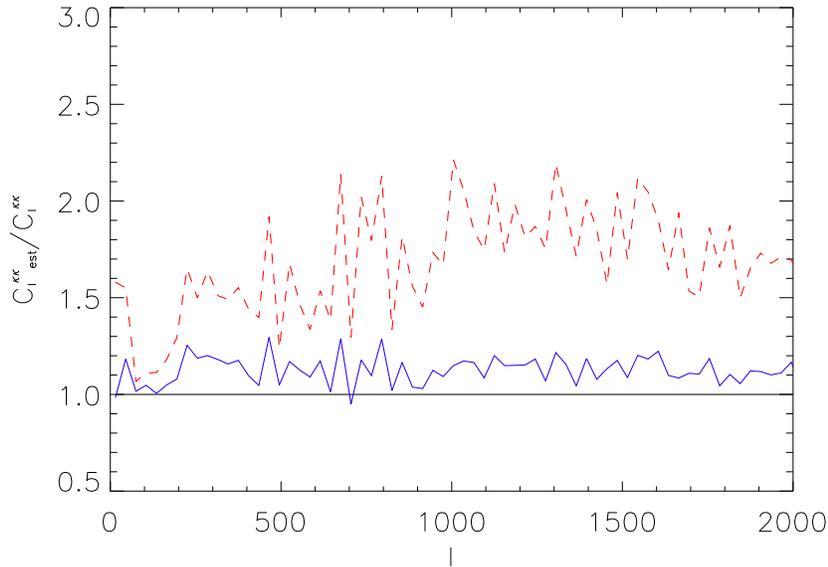}}
{\caption{As in Fig. \ref{fig:NGwSZspec}, but the kSZ has been masked, as 
described in the text.  The cross-spectrum (blue) is down to a bias 
of $\sim 15\%$, which is only slightly higher than when no kSZ is included.  
The bias in $C_{\ell}^{\kappa\kappa}$ (red) is reduced but still 
significant. }
\label{fig:Gspec3}}
\end{figure}

In any case, a small reduction in signal is an acceptable price to 
pay if we can control the large error introduced by the kSZ.  Indeed, 
the map made using this correction is greatly improved;  as we 
show in Fig. \ref{fig:NGwSZrec}, the spurious features have largely 
disappeared in the reconstruction, and the map 
made using masking is visually very similar to the reconstructed map 
made without including any kSZ at all.

The estimate of the power spectrum $C_{\ell}^{\rm est}$ also shows 
substantial improvement, and is now biased by only roughly $70\%$.  
Although this is considerably better than the unmasked case, there 
is obviously still a residual kSZ contribution of about $20\%$, which 
may arise from the kSZ which is uncorrelated to the lensing signal.  
There may also be a similar effect due to Ostriker-Vishniac fluctuations 
from $z>2$, but we have not studied this issue here.

It is clear that the kSZ is an important foreground which cannot safely 
be ignored when considering CMB lensing effects, and that the masking 
technique we have used here is a step forward.  We have not attempted to 
optimize the process, so it is not clear to what extent the masking 
procedure can be tuned to further reduce the contamination of the 
lensing signal.

\section{Noise and Bias for Upcoming Surveys} \label{sec:surveys}

In this section we shall model the reconstruction that may be 
possible from a number of upcoming experiments.  Our conclusions 
will be overly optimistic in that they ignore systematic effects 
and incomplete foreground removal, but they serve to illustrate the 
role of noise levels and resolution which can be achieved in the 
next 5 years.

\subsection{The Impact of Improving Angular Resolution} \label{sec:bias}

\begin{figure}[t]
\centerline{
\includegraphics[width=12cm]{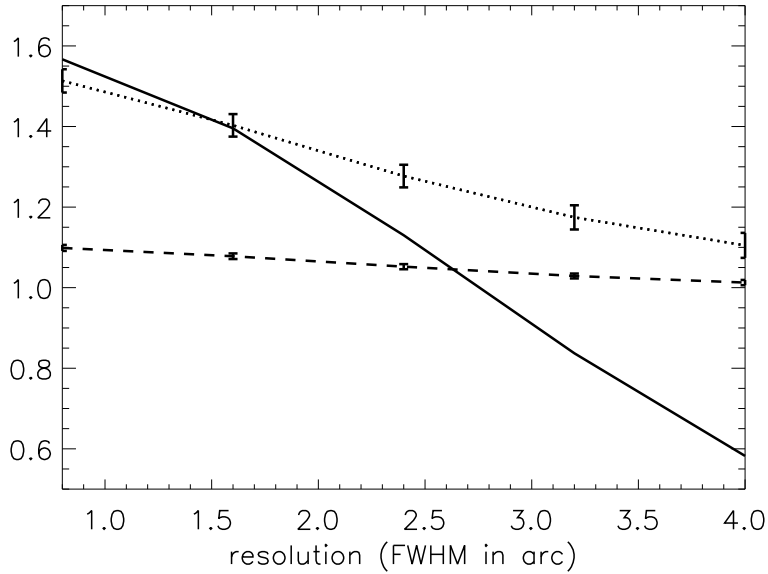}}
{\caption{Additive (dotted line) and multiplicative (dashed line)
average bias produced by the non-Gaussianity in the convergence field
versus the resolution of the experiment, the solid line represents the
average signal-to-noise ratio (square root of the signal variance over
the noise variance).  The averages are computed between $\ell = 0$ and
$\ell = 2000$.  The bias decrease with the resolution, but reducing
the bias lowers the sensitivity.}
\label{fig:biasres}}
\end{figure}

Until now, we have considered a high-sensitivity, high angular
resolution fiducial experiment, which has led us to find substantial
biasing effects related to the ability of the experiment to resolve
small structures.  In this section, we show the effect that improving 
the angular resolution has on the biases and on the signal-to-noise.
We compute the estimated convergence power-spectrum for different levels 
of angular resolution, keeping the noise level at a constant 
$2 \mu$K-arcmin, and find that both the
map bias (multiplicative bias $C_{\ell}^{\rm
cross}/C_{\ell}^{\kappa\kappa}$) and the noise estimate bias (additive
bias $C_{\ell}^{\rm est}/C_{\ell}^{\kappa\kappa}$) decrease as the
beam size increases (Fig. \ref{fig:biasres}).

\begin{figure}[h]
\begin{center}
\includegraphics[width=12cm]{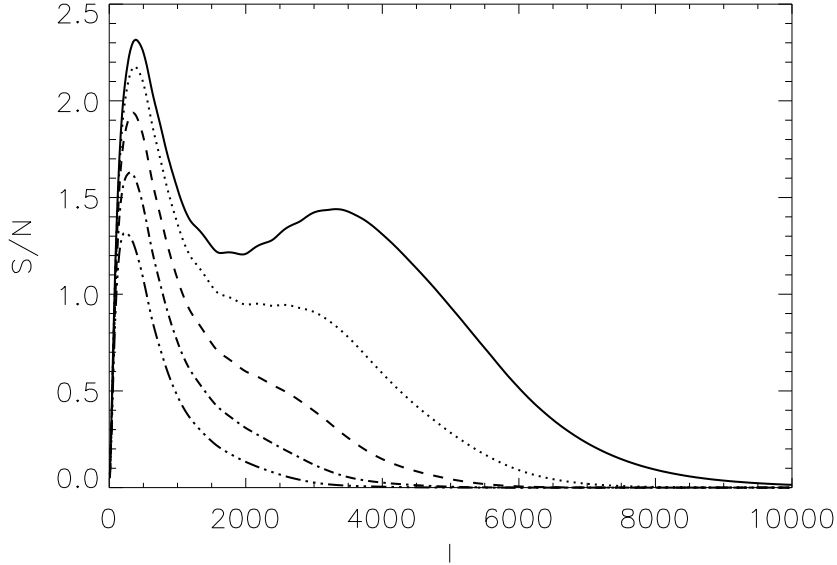}
\end{center}
{\caption{Signal-to-noise ratio versus angular scale for several
angular resolutions (respectively 0.8, 1.6, 2.4, 3.6, 4 arcminutes for
the solid, dotted, dashed, dot-dashed, triple dot-dashed lines). The
signal-to-noise ratio decreases with the angular resolution especially
at small angular scales.}
\label{fig:signoise}}
\end{figure}

The price of decreasing the bias is a much lower signal-to-noise
ratio (Fig. \ref{fig:biasres}), which is reduced on average by a
factor 3 when the resolution is degraded by a factor 5. Furthermore,
this lower angular resolution leads to a cutoff in $\ell$ space (Fig.
\ref{fig:signoise}; 4 arcminute resolution cuts the S/N at $\ell 
\sim 1000$) as the S/N is not constant in $\ell$. We got similar 
results by low-pass filtering the data of 0.8 arcminute beam (this
corresponds to change artificially the ``optimal'' filter), this easy
solution to reduce the bias degrades also the signal-to-noise
ratio. Adding the noise level as an another free parameter, one may
find an optimal (or several) values of the noise level and angular
resolution which minimize bias from non-Gaussianity in the convergence
field and maximize the average S/N. This preliminary study suggests
that a middle range angular resolution experiment (around 4 to 5
arcminutes), with a big sky coverage and a great sensitivity will be
best suited to accomplish this.

\subsection{Estimated Reconstructions for Upcoming Surveys} 
\label{sec:experiments}

\begin{figure}[h!]
\begin{center}
\includegraphics[width=12cm]{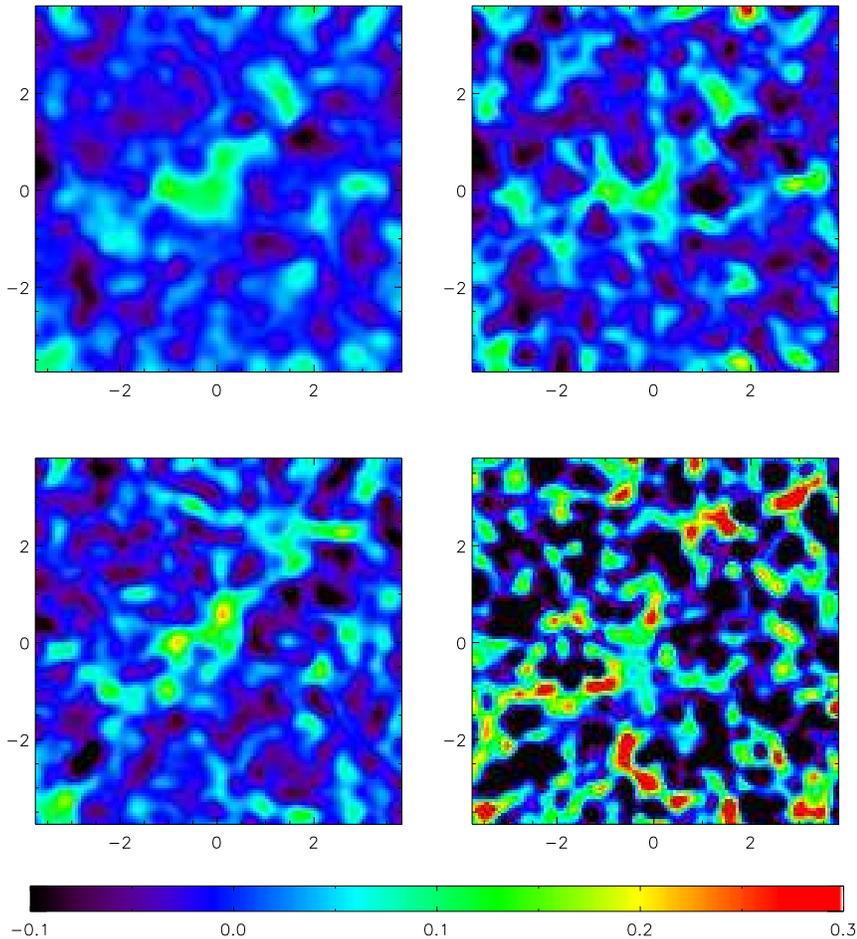}
\end{center}
\caption{Reconstructed maps of the convergence $\kappa$, from top to 
bottom and
left to right : input convergence map, APEX/SZ, SPT, Planck
reconstructed convergence map smoothed by a 20' FWHM Gaussian. The
maps represent a $7.5\times7.5$ degree chunk of each simulations. As
the noise level worsen the structures became less and less
recognizable. In Planck case, the map is completely dominated by noise
on these scales.}
\label{fig:apexspt}
\end{figure}

\begin{figure}[h!]
\centerline{
\includegraphics[width=8cm]{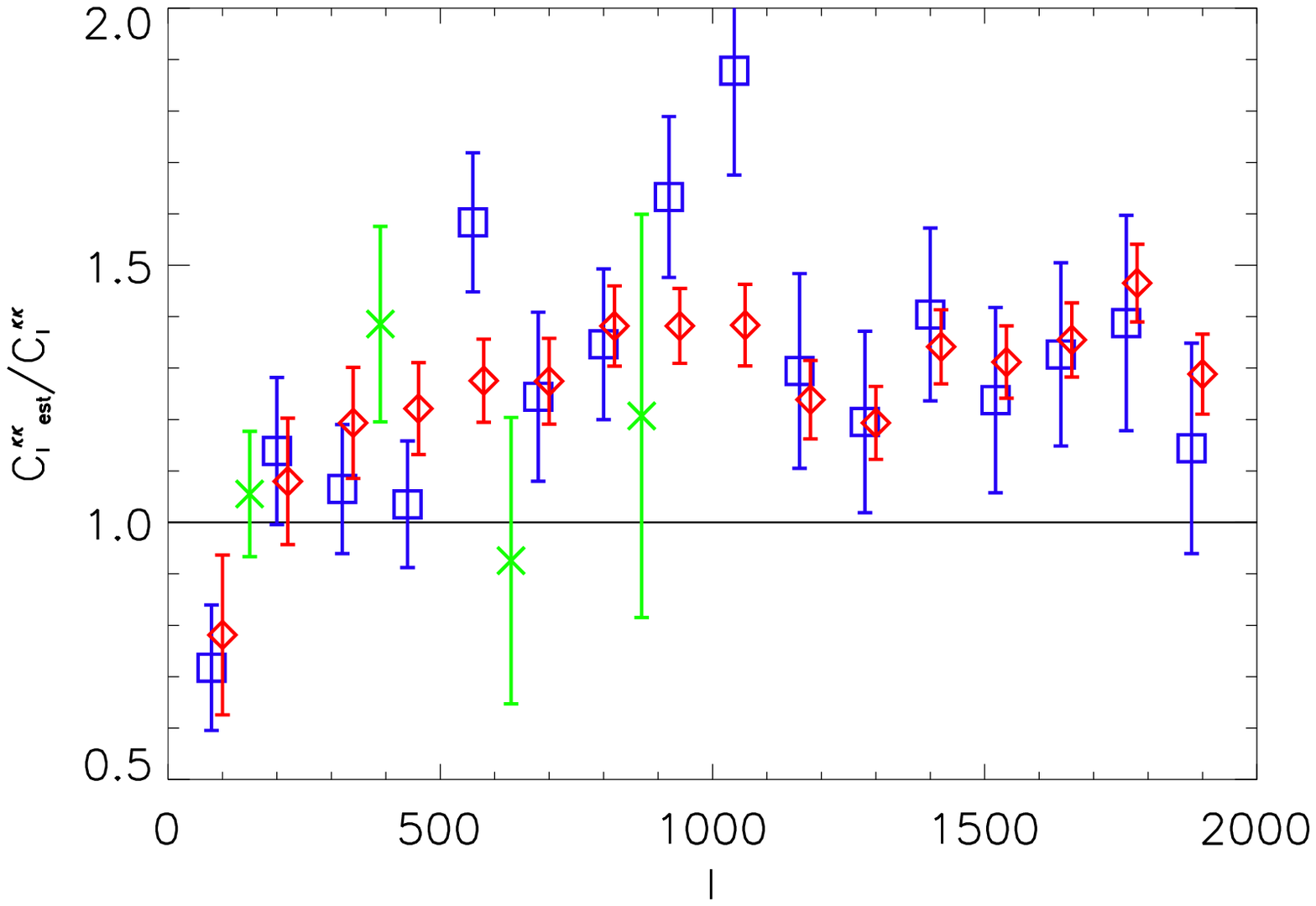}
\includegraphics[width=8cm]{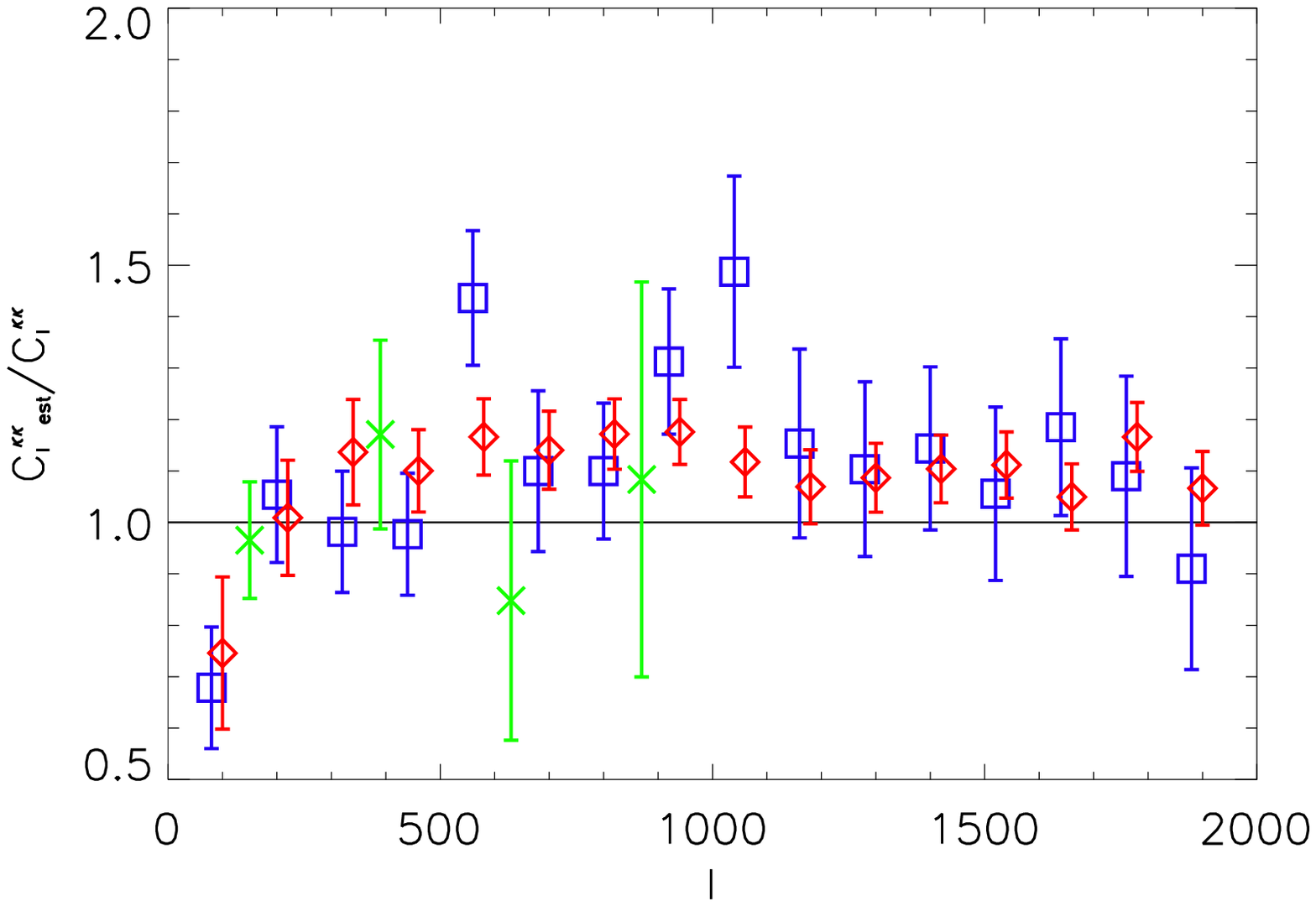}}
{\caption{Ratio of the estimated convergence power spectrum with the
input one for APEX/SZ (blue squares), SPT (red diamonds),
1/10$\rm^{th}$ of Planck (green crosses). The values are binned in
$\Delta\ell = 120$, except for Planck ones which are binned in twice
bigger $\Delta\ell$ (the Planck points stop at $\ell=1000$ as we get a
poor reconstruction on smaller scales).  The unmasked power spectrum
estimation (left) is biased on average by 30\%, whereas the masked
power spectrum estimation (right) is biased by only 10\%.  Note that
we have simulated only about one tenth of the sky, so that the error
bars on the Planck estimate appear larger than they should.}
\label{fig:bestcase}}
\end{figure}

In this section, we consider four upcoming surveys which possess enough 
sensitivity to supply a first detection of the CMB lensing
effect : APEX/SZ, ACT, SPT and Planck. We reconstruct the 
convergence from a CMB temperature map, which is made using 
a non-Gaussian lensing field, and includes contamination from the 
kSZ.  We include instrument effects appropriate to each experiment, and 
treat the kSZ in the maps using our ``masking'' correction technique.  
Small regions ($7.5^\circ \times 7.5^\circ$) of reconstructed 
$\kappa$ are displayed in Fig. \ref{fig:apexspt}, and the estimated power 
spectra are given in Fig. \ref{fig:bestcase}.

For APEX/SZ, we use a sensitivity of 8 $\mu$K-arcmin on a $15^{\circ}
\times 15^{\circ}$ field with a $0.^\prime 8$ angular resolution.
Note that these characteristics are sufficiently similar to those of
ACT (5.7 $\mu$K/arc$^2$ sensitivity on a $15^{\circ} \times
15^{\circ}$ field with a $1.^\prime1$ angular resolution) that we
refer the reader to the same plots for ACT as for APEX/SZ.  The
estimated convergence map is noisy, but the main features are
recognizable. Therefore APEX/SZ may well be the first experiment to
detect the CMB lensing effect. The surveys should also be sensitive
enough to place some limits on the power spectrum, although the
estimates will suffer from a non-negligible bias.

For the Planck satellite, we assume a 65 $\mu$K-arcmin sensitivity,
a $5^\prime$ resolution, and analyse only a $60^{\circ} \times 60^{\circ}$
field (about 1/10$\rm^{th}$ of Planck coverage) to remain in the flat
sky approximation. The reconstructed small scale convergence map is
completely dominated by noise, as expected and would also be true for a
full sky map. Using our masking technique the bias in Planck
reconstruction seems to be quite negligible, though our contrains on the
bias residual is limited by our small sky coverage. The full sky coverage
will reduce the error bars presented on Fig. \ref{fig:bestcase}, but the
convergence power spectrum estimate will be cut off at small scales
($\simeq$ 1000) due to the large size of the beam.

For SPT, we assume a 10 $\mu$K-arcmin sensitivity on a 
$60^{\circ} \times 60^{\circ}$ field with a $1^\prime$ angular resolution.  
The higher noise level makes the structures harder to recognize, although 
it is still possible to identify many of them.  The result for the power 
spectrum is markedly better than the other surveys we are considering;  
the combination of low noise, small beam, and large survey area combine 
to perform relatively well at this task.  

For the conditions we have assumed, all of these experiments will
detect the lensing effect and measure the convergence power spectrum
on scales ranging from a few degrees to a few arcminutes.  However,
although the masking technique shows promise for reducing this
contamination, even if the procedure was not optimized, the kSZ bias
remains significant for APEX/SZ, ACT and SPT. Our investigations
suggest that very low noise and large sky coverage are the two main
drivers for lensing reconstruction.

\section{Discussion} \label{sec:summary}

Lensing of CMB photons may provide a window into the matter distribution 
projected from primordial times to the present.  A great deal of work 
has been put into developing the statistical methods needed to fulfill 
this promise, and our goal here has been to further these efforts.  To this 
end, we have extensively checked the power and assumptions of one of the 
two principal estimators that have emerged, the ``optimal'' quadratic 
estimator of \cite{Hu01a}, both for reconstructing projected mass 
maps and for measuring the matter power spectrum.  Although our 
results apply specifically to the quadratic estimator, we note that 
the other principal candidate 
\citep[the maximum likelihood estimator of][]{Hirata03} makes the 
same assumptions about the underlying fields as the quadratic estimator, 
so it is likely to face the same challenges.

We began our efforts in Section \ref{sec:gauss}, where we tested the 
quadratic estimator method under the artificial conditions of 
a Gaussian gravitational potential, and uncorrelated Gaussian noise, 
for which the estimator is most likely to succeed.  The noise in the 
reconstructed maps has $\kappa$ dependent extra terms which constitute an
additive bias in power spectrum measurements.  
The second order terms were discussed in \citep{Cooray03}, and we 
confirm their prediction that these are numerically significant.  In 
addition, we find for the first time that higher order
terms are also likely to be significant, since these second order terms 
do not fully correct bias for interesting observational 
parameters.  Although the bias is signal dependent, it may be possible 
to correct for this fact by using iterative 
methods or model fits, as suggested by \cite{Kesden03};  however, high 
order terms would need to be included to obtain good accuracy.

It is clear that a reconstruction method based on the assumption 
of a Gaussian lensing field will not optimally treat non-Gaussian 
features, such as clusters, which contain more information than just 
the 2-point function.  We investigated this issue in Section 
\ref{sec:nongauss} by using a lensing field generated from an 
N-body simulation.  This imposed both a multiplicative bias on the
cross spectrum estimate, and an additional additive bias on the auto
spectrum.  The bias increased as the signal-to-noise ratio increased.

We investigated the impact of a second non-Gaussian complication, the 
kinetic Sunyaev-Zel'dovich effect (kSZ), in Section \ref{sec:ksz}.  
This ``noise'' is highly correlated with the lensing signal and, unlike 
the larger thermal SZ, is impossible to distinguish from the CMB using 
spectroscopic methods.  We began by showing that, if left untreated, 
this effect will completely dominate the reconstruction.  
One obvious method to correct for this is to use a measurement 
of the thermal SZ (which is correlated with the kSZ) to mask pixels 
which are likely to have a large contamination, and we were 
able to substantially reduce the bias caused by the kSZ 
while excising only a relatively small number of pixels.  
Although this was encouraging, not all of the kSZ signal is correlated
with the thermal SZ and the contamination level remained significant.
Further reduction of kSZ noise by simply masking more pixels suffers from
steadily decreasing returns in signal-to-noise improvement.
We suspect, although we have not checked it explicitly, that contamination
{}from the Ostriker-Vishniac effect would behave in a similar manner.

Given the difficulty that one encounters using Gaussian reconstruction 
methods, it is natural to ask if the non-Gaussian nature of the 
lensing signal might itself provide the best probe of the lensing 
field.  This topic has recently been the subject of great interest 
in the context of galaxy lensing \citep[e.g.][]{TaJa03,ZaSc03,HoWh03}, 
but to this point a similar effort is lacking for lensing of the 
CMB.  There is clearly promise in this idea; however, any measurement of 
the non-Gaussian signal is likely to be particularly sensitive to 
the masking technique needed to control the kSZ, which will be highly
correlated with this signal.  
For example, clusters will produce both a large non-Gaussian lensing 
signal and a large kSZ in the same location, \emph{at the cluster}.  
Thus, it may be necessary to mask precisely those pixels which would 
otherwise have provided the best window into the non-Gaussian signal.

In this paper, we have been primarily interested in the 
reconstructions that can be accomplished using the unprecedented 
power and resolution that will be available in the next generation 
of surveys, and it is in this context that biasing effects 
emerge as important obstacles.  In the regime of large smoothing 
and low signal-to-noise, the bias due to non-Gaussian effects is 
small, and the quadratic estimator performs well in this regime.  
It is only when the observational parameters are improved that the 
biasing effects we have been discussing emerge.  We show 
in Fig. \ref{fig:biasres} the unfortunate increase in bias that 
occurs as signal-to-noise is improved, so that a recovery of 
the power spectrum can have either small error bars or negligible 
bias, but not both.  Thus, while a detection of the lensing effect 
is likely to be achieved by APEX/SZ, more ambitious goals 
that rely on extremely high fidelity reconstructions of the matter 
power spectrum require significantly more modeling.

We have presented a number of reconstructed maps in this 
paper, and it is clear that features in these maps do coincide 
with the those of the original fields at some level.  However, 
we note that the maps have been smoothed by a $20^\prime$ 
FWHM Gaussian window, and even so the reconstructed maps are 
far from perfect.  We have also performed 
extensive numerical integrations to compute the corrections to 
the noise term $\mathcal{N}_{\ell}$ outlined in Section \ref{sec:gauss}.
Although these corrections proved useful for for removing some of the 
bias effects, the computational power needed for this calculation 
is non-trivial, a point which we discuss in more detail in the 
Appendix.  

Although we have not treated the issue here, we would like to make 
some comments about the role of polarization in CMB lensing.  It has 
been emphasized \citep{Guzik00,Hu02,Cooray03,Hirata03b,Kesden03,Okamoto03}
that the inclusion of polarization information might dramatically 
enhance the prospects for large-scale structure reconstruction from 
lensing of the CMB.  This is because lensing induces a $B$-mode 
polarization signal which is otherwise absent for purely scalar, 
primary fluctuations.  The large intrinsic primary CMB anisotropies, 
which are a source of ``noise'' for lensing reconstruction, are 
thus absent.  However, the spatial structure is complicated for 
polarization, as it is for temperature, and the signal levels are 
much smaller.  The kSZ effect, which is one of our major contaminants, 
is also polarized \citep{SZ80b}.  Thus, although it is certainly 
reasonable that the addition of polarization information would enhance 
the prospects for $C_\ell^{\kappa\kappa}$ reconstruction, a detailed 
calculation is required to determine how much better one can actually do.

Weak lensing is one of the principal tools for cosmologists, and promises 
to be increasingly significant for many years to come.  Lensing of 
the CMB is a new addition to the arsenal, which will be facilitated by 
powerful upcoming surveys such as ACT, APEX-SZ, Planck, and SPT.
Even as observers gear up to probe the millimeter and sub-millimeter 
wavebands with unprecedented power and resolution over large fractions 
of the sky, theorists continue to improve the methods necessary to 
extract relevant information from the resulting measurements.  If the 
pace of development on both fronts continues at its current rate, the 
future looks bright indeed.

\noindent {\bf Acknowledgments:}\newline
A.A. would like to thank the organizers of the workshop ``Cosmology with
Sunyaev-Zel'dovich cluster surveys'' held in Chicago in September 2003 for
allowing him the chance to present some of this work.  Additionally we would 
like to thank T. Chang, J. Cohn, M. Zaldarriaga for helpful discussions
about these results.
The simulations used here were performed on the IBM-SP2 at the National
Energy Research Scientific Computing Center.
This research was supported by the NSF and NASA.

\appendix \section{Appendix: Numerical Issues} \label{Appendix}
\subsection{Computing the Quadratic Estimator} \label{Nell}

In this section, we discuss some of the issues associated with computing 
the quadratic estimator of \cite{Hu01a} and the higher order corrections 
we have considered in this paper.  One of the advantages of the estimator 
is that it is inexpensive to compute, but higher order terms add to the 
difficulty, as we discuss below.

The lensing effect can be expressed as a Taylor expansion 
(Eq. \ref{eq:linear}) of the unlensed temperature.  If the
deflection angle $\delta \thetab$ is small enough, then the
linear approximation is valid, and the normalization 
$A_\ell$ (Eq. \ref{eq:N_L}) of the optimal quadratic estimator 
(Eq. \ref{eq:hustimator}) is a good approximation to the
estimated convergence power spectrum noise term.
In this case, the estimator requires as inputs the measured CMB 
temperature map and the power spectrum of the unlensed CMB.  It is 
likely that a good estimation of the latter can be achieved by measuring 
cosmological parameters at large angular scales.

\cite{Cooray03} explored terms to second order in the deflection angle.  
Although this resulted in the same estimator of $\kappa$, they found another 
significant term in the estimator $C_{\ell}^{\rm est}$ of the 
convergence power spectrum, in addition to the first order noise 
correction $A_\ell$.  This term, $A_{\ell}^{NG}$, is defined as  
\begin{eqnarray} \label{eq:ng}
&& \small{
A_\ell^{NG}={A_\ell^2\over \ell^2}\int{d^2\ell_1 \over (2\pi)^2}
\int{d^2\ell_2 \over
(2\pi)^2}F(\ellb_1,\ellb-\ellb_1)
F(\ellb_2,\ellb-\ellb_2)}\\
&& \nonumber\footnotesize{
\left\{C^{\phi\phi}_{|\ellb_1-\ellb_2|}
f(-\ellb_1,\ellb_2)f(\ellb_1-\ellb,\ellb-\ellb_2)
+C^{\phi\phi}_{|\ellb_1-\ellb+\ellb_2|}
f(-\ellb_1,\ellb-\ellb_2)f(\ellb_1-\ellb,\ellb_2)\right\}}
\end{eqnarray}
where $F(\ellb_1,\ellb_2)$ is as in Eq. \ref{eq:filter} and 
$f(\ellb_1,\ellb_2)= {(\ellb_1+\ellb_2) \cdot (\ellb_1 \tilde{C_{\ell_1}} +
\ellb_2 \tilde{C_{\ell_2}})}$.
This term is smaller than the $0^{\rm th}$ order term $A_\ell$, 
but it is not negligible;  for example, it represented $\sim 35\%$ of 
the power for our fiducial low noise, high resolution survey used in 
Sections \ref{sec:gauss} and \ref{sec:tests}.  The convergence of this 
integral and the one of equation \ref{eq:N_L} is provided by the noise 
level, so that the computation time increases as lower noise surveys 
are considered.  However, the second order terms are not a problem 
for a modern workstation, and takes about 30 minutes of CPU time for 
a typical calculation.

The computation of Eq. \ref{eq:ng} introduces a new challenge, in 
that this requires the potential power spectrum $C_{\ell}^{\phi\phi}$ as 
an input, which is of course the quantity we are 
trying to measure.  Either an iterative procedure or a model fit might 
resolve this issue, and we have assumed (but not demonstrated) that 
this is possible, so that we use the actual input $C_{\ell}^{\phi\phi}$ for 
our computation of $A_{\ell}^{NG}$.

When we do this, 
the inclusion of $A_{\ell}^{NG}$ into the estimate of the 
power spectrum improves the bias issue substantially, but does not 
solve the problem.  Additional terms may be required, so that the 
auto-spectrum of the map may include other terms, written schematically 
below as 

\begin{equation}\label{eq:schetayexp}
C_{\ell}^{\rm auto} \simeq \left< T^4 \left(a + b\kappa + c\kappa^2 + 
d\kappa^3 + e\kappa^4 + \cdots + i\kappa^8\right)\right>
\end{equation}

So far, by including $A_\ell$ and $A_{\ell}^{NG}$, we have considered 
the 0$^{\rm th}$, 2$^{\rm nd}$, and a part 4$^{\rm th}$ order terms 
in $\kappa$.  Some of the terms we have ignored may be small enough to be
irrelevant, and in the case of a Gaussian $\kappa$, the odd order
terms should be null. However, our results in section \ref{sec:gauss} 
exhibit evidence for a non-negligible contribution from  additional 
higher order terms.  If $\kappa$ is a Gaussian field, 
$C_{\ell}^{\rm auto}$ can be written analytically, and in principle its 
noise contribution computed.  
However, the numerical challenge would be non-trivial.  Since the $\kappa^2$ 
correction took $\sim 30$ minutes, we estimate that for the same
resolution, the $\kappa^4$ would take 5,000 hours, and so on for higher
orders.  Furthermore, if $\kappa$ is a non-Gaussian field, then the odd 
order moments are not expected to be identically zero, and the higher 
order even moments can not be expressed as a function of the two point 
correlation function.  In this case, numerical simulations will be needed 
to recover the convergence power spectrum.

\subsection{The Simulations} \label{simulations}

In this paper, we have made extensive use of maps of the primary CMB 
temperature anisotropies, the convergence $\kappa$, and the thermal 
and kinetic SZ effects.  Our technique for simulating these maps has been 
described in detail elsewhere \citep{VAW04}, so we confine 
ourselves here to a brief discussion of some points directly 
relevant to this work.  

We chose a fiducial flat, $\Lambda$CDM cosmology generally consistent 
with currently popular models, with $\Omega_m=0.3, \Omega_{\Lambda}=0.7, 
h=0.7, \Omega_bh^2=0.02, n=1,\sigma_8=1$.  Maps of primary CMB 
temperature anisotropies were made as random realizations of a Gaussian 
field with a power spectrum computed using CMBfast \citep{SelZal96}.  The 
CMB maps are $30^{\circ} \times 30^{\circ}$ square patches of sky 
containing $2048 \times 2048$ pixels.

We made use of Gaussian convergence maps to account for the lensing 
effect of matter located at high redshift $z > 2$, and for certain 
tests as outlined in the text.  These were created from a random 
realization of a Gaussian field in much the same manner as the maps 
of the CMB, but using a power spectrum computed using the the non-linear 
growth factor described by \cite{PD96} and the Limber approximation. 

The other maps require some knowledge of the spatial distribution 
and time evolution of the mass in the model, which we obtain from 
an N-body simulation using the TreePM code described in 
\cite{CodePaper}.  This simulation modeled a large volume of the 
universe, a cube $300 \ h^{-1}\rm Mpc$ on a side, and used only 
dark matter, which was modeled using $512^3$ particles.  This was then 
used to make maps of the convergence, the thermal SZ, and the kinetic 
SZ, for a distribution of matter out to a redshift $z = 2$ 
and a field of view of $7.5^{\circ} \times 7.5^{\circ}$.  The 
convergence map was made by a simple weighted sum of the line of 
sight density contrast and thus implicitly assumed the validity of the 
Born and Limber approximations 
\citep[which was investigated in some detail in][]{VW03}.  
In order to compute the thermal and 
kinetic SZ effects, we assumed that the baryonic matter traces the dark 
matter, which is likely to be a good approximation on scales 
which can be resolved by our fiducial surveys.  We refer the reader to 
\citep{ShWh03,VAW04} for a complete description of this process.

\begin{figure}[h]
\centerline{
\includegraphics[width=\textwidth]{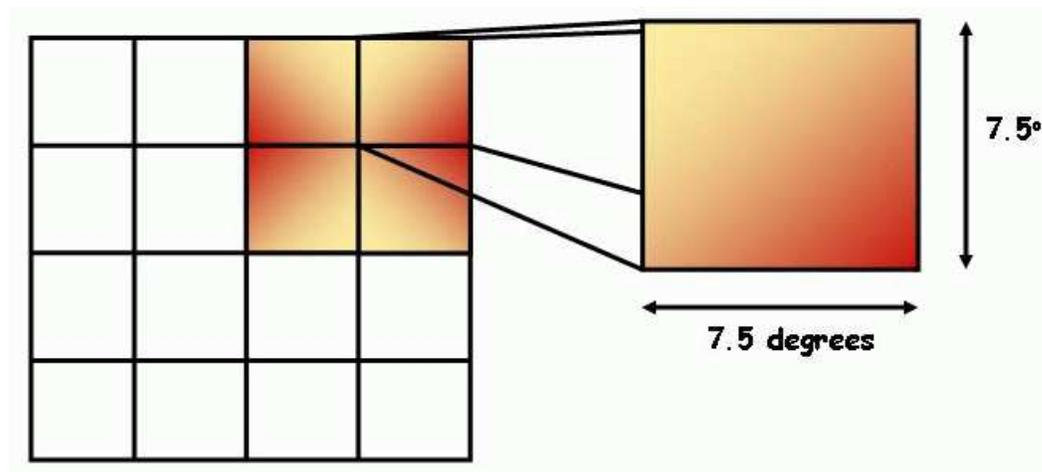}}
\caption{An illustration of the tiling technique we use to create 
a large ($30^\circ \times 30^\circ$) map from a smaller one.  Maps 
of the convergence $\kappa$ and of the SZ effects are derived using 
the matter distribution from an N-body simulation with a 
$7.5^\circ \times 7.5^\circ$ field of view.  To avoid artificial 
boundaries in the larger field, we cover it with sixteen tiles.  
The tiles are identical to one another and are oriented in a manner 
that assures a continuous field with periodic boundary conditions. }
\label{fig:flip}
\end{figure}

The reader may have noticed that the CMB temperature maps are $30^{\circ}$ 
on a side, while maps made using the N-body simulation are only 
$7.5^{\circ}$ on a side.  The reason for this discrepancy is that 
simulations that have the dynamic range to cover so large a field of 
view while maintaining sufficient small scale resolution require more 
computing power than we can muster.  We could cover the field with 
patches made from different simulations, but this would create highly 
artificial boundaries.  Our solution to this issue, as illustrated in 
Fig. \ref{fig:flip}, is to cover the $30^{\circ} \times 30^{\circ}$ patch 
of sky with sixteen $7.5^{\circ} \times 7.5^{\circ}$ tiles of a given 
field.  These smaller patches are duplicates of the same field, but are 
oriented so that the resulting map avoids the structural discontinuities 
mentioned above.  In this way we are able to create the large continuous 
periodic fields necessary to test reconstruction techniques.

\end{document}